\pdfoutput=1
\documentclass[preprint,twocolumn,tightenlines,10pt,prl]{revtex4}%
\usepackage{amsfonts}
\usepackage{amsmath}
\usepackage{amssymb}
\usepackage{graphicx}%
\providecommand{\U}[1]{\protect\rule{.1in}{.1in}}

\pdfoutput=1
\begin{document}
\title{ Wireless Josephson Junction Arrays as Tunable Metamaterials: \\Inducing Discrete Frequency Steps with Microwave Radiation}
\author{L. L. A. Adams}
\affiliation{School of Engineering and Applied Sciences, Department of Physics, Harvard
University, Cambridge, MA 02138.}

\pacs{61.46.Df, 61.46.Hk}

\begin{abstract}
We report low temperature, microwave transmission measurements on a new
switchable and tunable class of nonlinear metamaterials. A wireless two
dimensional array of Josephson junctions (JJ) is probed as a metamaterial
where each plaquette in the array is considered as a meta-atom. In the
presence of microwaves, this compact metamaterial of 30,000 connected
meta-atoms synchronizes the flow of Cooper pairs to yield a single robust
resonant signal with a quality factor of 2800 at the lowest temperature for
our measurements. The transmission signal is switched on and off and its
amplitude and frequency are tuned with either temperature, incident rf power
or dc magnetic field. Surprisingly, increasing the incident rf power above a
threshold causes the resonance to split into multiple discrete resonances that
extend over a range of 240 MHz for a wide temperature window. We posit that
this effect is a new incarnation of the inverse ac Josephson effect where the
Josephson plasma frequency locks to the rf drive frequency generating
quantized frequency steps, instead of voltage steps, in the absence of a dc
bias. \TeX{} .

\end{abstract}
\volumeyear{year}
\volumenumber{number}
\issuenumber{number}
\received[Received text]{date}

\revised[Revised text]{}

\startpage{1}
\endpage{102}
\maketitle

Prompted by theory \cite{veselago},\cite{pendry}, extensive studies of
carefully designed artificial materials, known as metamaterials, are underway
to find new electromagnetic phenomena such as amplification of evanescent
waves \cite{liu} - \cite{adams}, extraordinary transmission \cite{Barnes},
\cite{Tsiatmas}, negative index of refraction \cite{pendry2}, \cite{smith},
and classical analogues of electromagnetically-induced transparency
\cite{zhang} - \cite{Gu}. However, in spite of recent advances using metallic
metamaterials, requisites for practical applications such as switching on and
off a signal and modulating a signal's frequency and amplitude using external
control parameters still need to be addressed. Since fulfilling these
conditions with existing \textit{metallic} structures is challenging, because
of absorption losses, developing a \textit{new} class of \textit{tunable
}metamaterials, such as superconducting metamaterials, is arguably warranted.
\cite{anlage2} - \cite{Kawayama}

In this Letter, we show that tunable metamaterials already exist in the form
of superconducting Josephson junction, JJ, arrays. Most remarkable, however,
isn't their tunability\textit{ per se}, but discovering an unanticipated rf
power and temperature dependent manifestation of discrete frequency steps
that, to our knowledge, has not been previously reported. This physical
phenomenon opens up new possibilities for signal processing if scaled to THz
frequencies \cite{Lukin}, new opportunities for high frequency detectors for
radio telescopes \cite{Tucker}, and raises new theoretical questions about the
relevance of the inverse ac Josephson effect to microwave transmission
experiments. The inverse ac Josephson effect is the reverse of the ac
Josephson effect. In the ac Josephson effect, an applied dc voltage across a
JJ generates a periodically oscillating supercurrent, pairs of electrons, that
emits electromagnetic radiation whereas in the inverse ac Josephson effect,
shining electromagnetic radiation onto an unbiased JJ induces a dc voltage
difference across the tunneling barrier. Here we show that shining microwaves
on a wireless Josephson junction array induces seven discrete frequency steps
instead of voltage steps, and in between these resonances, there are narrow
opaque windows in the transmission line. The inverse ac Josephson effect only
applies for underdamped JJs \cite{Levinsen} which also describes the JJs
studied here.%
\begin{figure}
[ptb]
\begin{center}
\includegraphics[
height=2.565in,
width=3.4497in
]%
{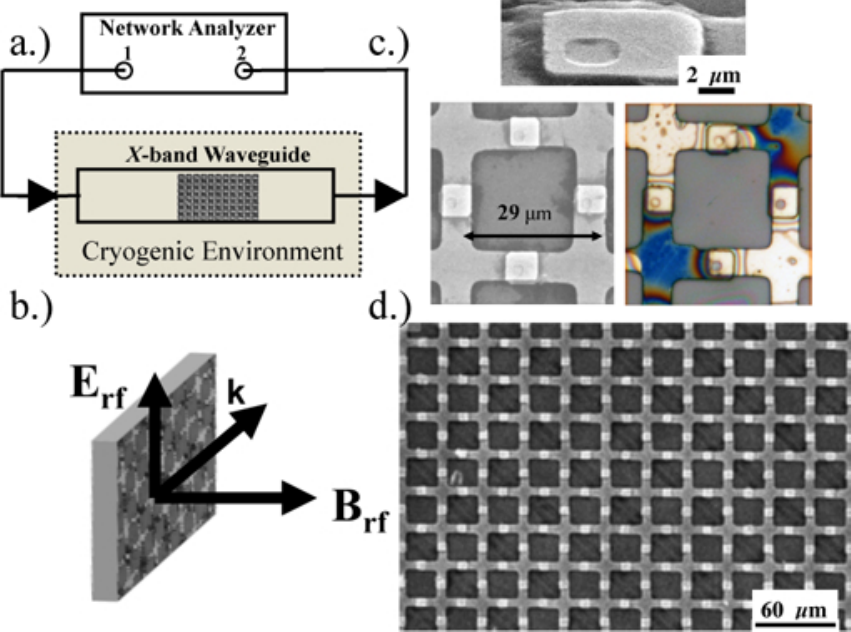}%
\caption{(Color online) Experimental set-up and sample a.) Schematic of the
experimental setup containing a network analyzer and a waveguide. b.)
Orientation of sample to electromagnetic waves in waveguide. c.) SEM\ images
of the JJ meta-atom: above, image of the tunnel junction which is shown in the
circular indentation in the image and below, two images of a single
JJ\ meta-atom with 4 JJ per loop, left image, in black and white, and right
image, in color. d.) SEM\ image of the JJ array showing only a small fraction,
\symbol{126}80 meta-atoms, of the sample.}%
\end{center}
\end{figure}

Two dimensional Josephson junction arrays are macroscopic quantum materials
with a highly tunable superconducting current. We consider a \textit{wireless}
two dimensional network of 30,000 \textit{connected }Josephson junction (JJ)
plaquettes with each plaquette having a square loop geometry containing 4
equally spaced JJs; we consider these plaquettes as meta-atoms which form, in
aggregate, a switchable and tunable metamaterial. In contrast to recent
microwave transmission reports on \textit{discrete} split ring JJ\ meta-atoms
\cite{ustinov} , \cite{anlage} , we find a sharp, robust resonance. As with
our previous measurements on this sample, abet at evanescent frequencies
\cite{adams}, the resonance is clearly distinct from noise without processing
the data or using any additional microwave components, such as amplifiers,
attenuators, directional couplers or filters, which also differentiates our
work from others.

The experimental apparatus, described in detail elsewhere \cite{adams},
consists of a vector network analyzer connected to a X-band waveguide via
phase-maintaining co-axial cables as seen in the schematic of Figure 1 a. The
waveguide, enclosed in a vacuum can, is cooled to cryogenic temperatures; it
is magnetically shielded by $\mu-$ metal to reduce the ambient field. The
JJ\ metamaterial is positioned in the center of the waveguide and oriented
such that the rf B-field is normal to the plane of the array as seen in the
schematic in Figure 1 b.

The sample is a portion of a much larger array \cite{kroger}, \cite{bhushaan};
its transport properties, before removing electrical contact pads and reducing
its size,\textit{ }have been extensively studied elsewhere. \cite{goldman}
Scanning electron microscope, SEM, images of a single tunneling junction,
Nb-amphorous silicon-Nb, and of a single meta-atom are seen in the upper and
lower left images, respectively, of Fig. 1 c. The 500 nm niobium thick square
loop is 29 $\mu m\times29$ $\mu m$ with a linewidth of 5 $\mu m.$ Each of the
4 JJs in the loop has an area of 2.5 $\mu m^{2}$ with an insulating barrier of
10 nm. \ A colored optical microscope image of a single meta-atom showing
niobium coverage is seen in the lower right image of Fig. 1 c; the
non-continous coverage of niobium indicates that resonance occurs because
charged particles must tunnel across capacitive tunneling barriers to
circulate throughout the network. A portion of the JJ metamaterial is seen in
the SEM image of Fig. 1 d. The individual JJ meta-atoms are about 1500 times
smaller than the microwave wavelength, which is approximately 40 mm, placing
them in the deep sub-wavelength limit.\cite{benz} The superconducting
transition temperature, T$_{c}$, of the individual niobium islands is 9.2 K.
\cite{squid}

If we consider that each JJ contributes a capacitance, C, estimated to be 0.9
pF, a normal state resistance, R, estimated to be 510 $\Omega,$ and a critical
current, \textit{I}$_{c}$, which is estimated to be 0.165 $\mu$A,
\cite{values}, we can estimate the degree of damping in a junction, using the
dimensionless McCumber parameter, $\beta_{c}$, which is $\beta_{c}%
=\frac{2eI_{c}R^{2}C}{\hbar}\!\thicksim118$, where \textit{e} is the electron
charge and $\hbar$ is Planck's constant divided by 2$\pi$. \ Since $\beta
_{c}>1,$ the junction is said to be underdamped meaning that there are
inertial effects of the "phase" particle \cite{newrock} which satisfy one of
the conditions required for the inverse ac Josephson effect \cite{Levinsen}. A
second condition that needs to be met is that the driving rf frequency should
be significantly larger than the natural Josephson plasma frequency,
$f\!=\frac{1}{2\pi}\!\sqrt{\frac{2eI_{c}}{\hbar C}}\thicksim$ 15 GHz.
\cite{Niemeyer} The resonant frequency, which is the plasma frequency
\cite{adams}, also depends, in part, on the sample's location in the waveguide
indicating the need for additional capacitive terms \cite{Devoret}, which
would lower the resonant frequency but not by enough to satisfy this second
condition. Nevertheless, we will show that the mechanism responsible for the
transmission features seen here are due to the inverse ac Josephson effect.\
\begin{figure}
[ptb]
\begin{center}
\includegraphics[
height=5.3333in,
width=3.0018in
]%
{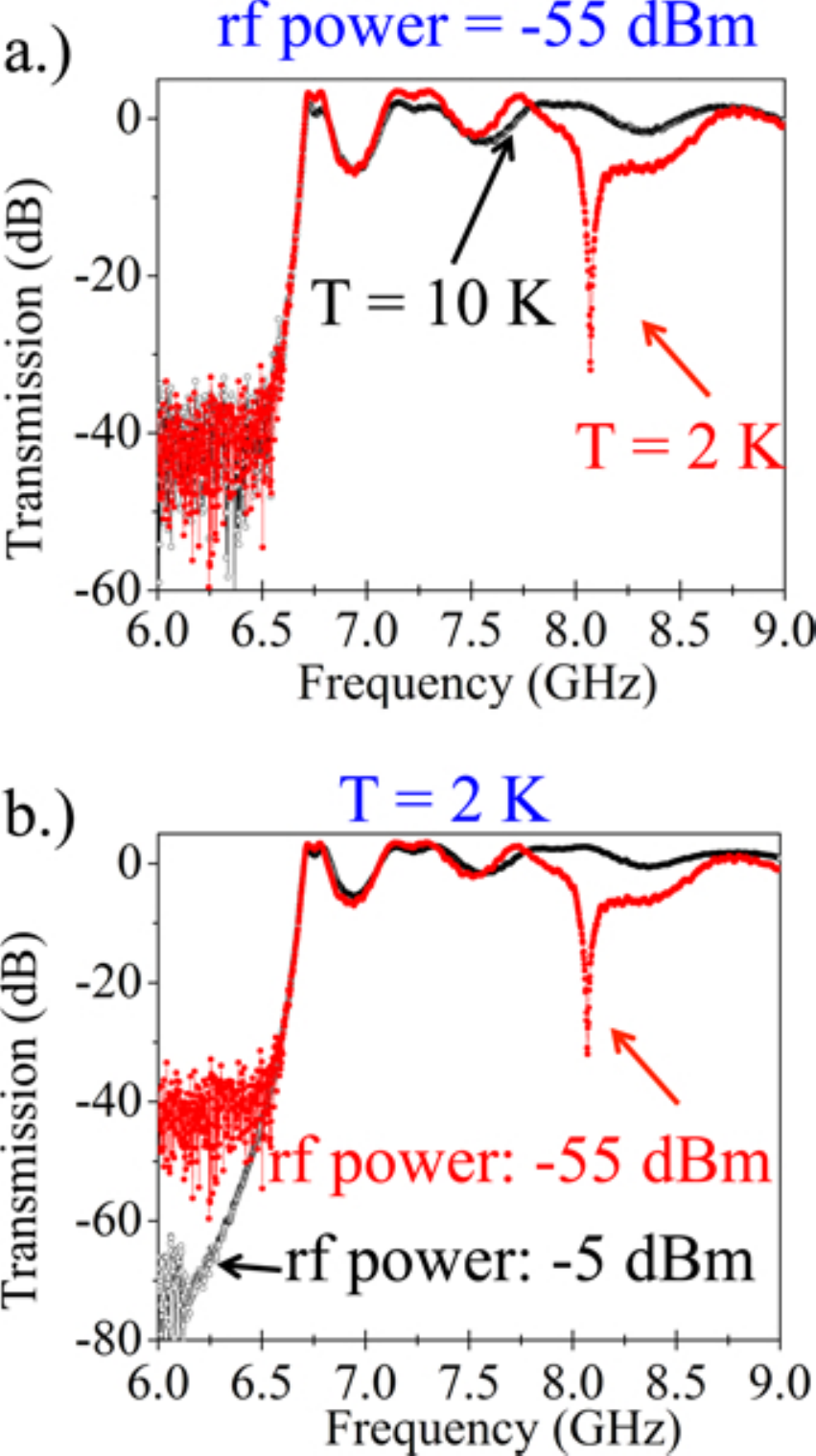}%
\caption{(Color online) Switching a resonance on and off with temperature and
rf power. a.) Temperature dependence at fixed rf power, -55 dBm, for T = 2 K
(red) and 10 K (black) and b.) rf power dependence at fixed temperature, 2 K,
for rf power -55 dBm (red) and -5 dBm (black). Note that the range in
transmission on the vertical axis is different in \textit{a }and \textit{b.}
Also note that the transmission values above 0 dB correspond to 0 dB: this
offset is due to the calibration being done at room temperature while the
measurements are at liquid helium temperatures.}%
\end{center}
\end{figure}
\qquad\qquad

Cooling the sample to temperatures below the superconducting transition
temperature and applying low power microwaves to the JJ array results in a
sharp resonance at \textit{f} = 8.08 GHz appearing in transmission as seen in
the plots of Fig. 2. Extinguishing the resonance is achieved by either
increasing the temperature above the superconducting transition temperature as
seen in Fig. 2 a, or by increasing input power as seen in Fig. 2 b.
\begin{figure}
[ptb]
\begin{center}
\includegraphics[
height=3.1375in,
width=3.5051in
]%
{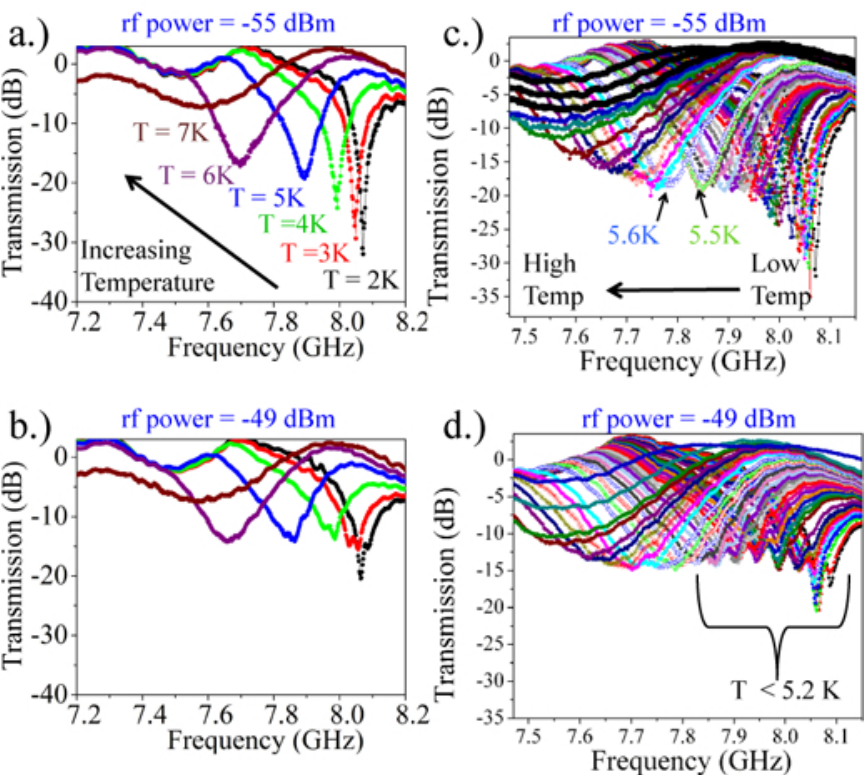}%
\caption{(Color online) Tuning the resonance with temperature and rf power.
Temperature dependence when the rf power is fixed at a. )-55 dBm and b.) -49
dBm for temperatures: T = 2.0 K(black), T = 3.0 K (red), T \ 4.0 K(green), T =
5.0 K (blue)= 6.0 K (purple), and T = 7.0 K (brown). Aggregation of
temperature-dependent data at fixed power: c.) -55 dBm and d.) - 49 dBm. Each
color represents a different temperature.}%
\end{center}
\end{figure}

Increasing temperature, at fixed rf power, shifts the resonance to lower
frequencies and broadens the resonance. This is seen in the transmission vs
frequency plots for different temperatures in Fig 3 a-d. As temperature
increases, Cooper pairs break lowering the critical current, I$_{c},$ and thus
shifting the resonance to lower frequencies. \ While the density of Cooper
pairs decreases, the density of normal conduction electrons increases, leading
to more dissipation and insertion loss; for instance, a jump of - 5 dB in
insertion loss is measured when the temperature reaches 7 K as seen at 7.2 GHz
in Fig. 3a and 3b. While these behaviors are reminiscent of those seen with
superconducting metamaterials \cite{ricci} , \cite{cihan}, they are not
identical. Here, the resonances are more sensitive to temperature increases
due to Cooper pairs tunneling across barriers as described by the
Ambegaokar-Baratoff current-temperature relation. \cite{Tinkham}%
\begin{figure}
[ptb]
\begin{center}
\includegraphics[
height=2.9836in,
width=3.4497in
]%
{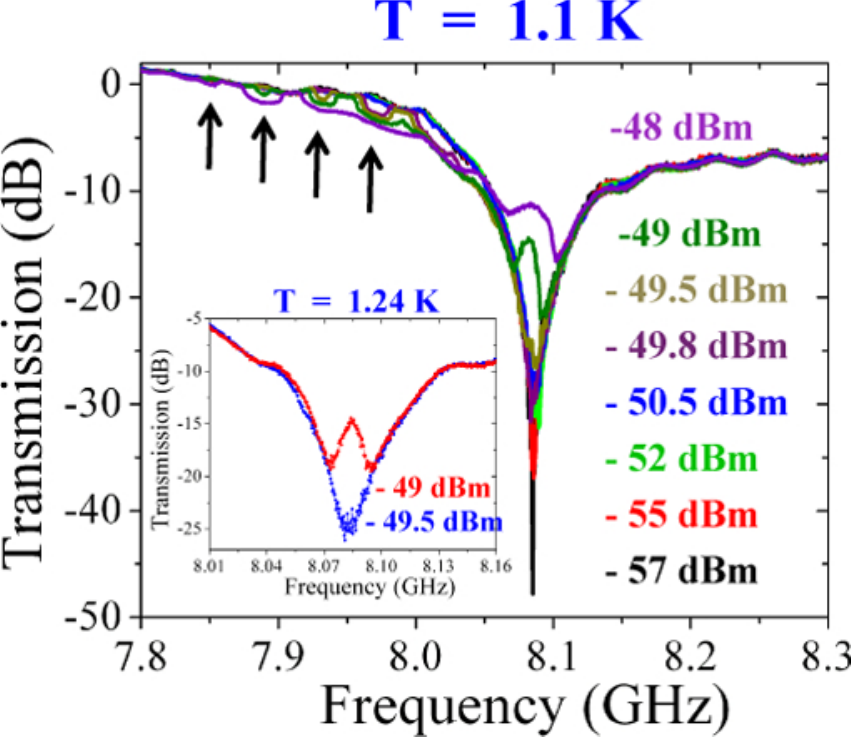}%
\caption{ (Color online) RF power dependence at fixed temperature, T=1.1 K.
The black arrows indicate multiple discrete resonances that appear below the
central resonance and at high rf power; these are spaced about 0.04 GHz apart.
Inset: RF power dependence at fixed temperature, T = 1.24 K for -49.5 dBm
(blue) and -49 dBm (red).}%
\end{center}
\end{figure}

Similar temperature dependent resonant responses persist when comparing
responses at two different rf powers, -55 dBm, as seen in Fig. 3 a, and -49
dBm, as seen in Fig. 3b. At -49 dBm, the higher rf power, the resonance also
shifts to lower frequencies while broadening with increasing temperature;
however, the overall amplitude decreases and the resonance dip splits into two
for T
$<$
6\ K. \qquad\qquad\ \ 

To elucidate the effect of the resonances splitting at -49 dBm, we compare two
large sets of temperature data, one at -55 dBm and the other at -49 dBm as
seen in Fig. 3 c and 3 d. The aggregate temperature data at -49 dBm and for T
$<$
5.2 K clearly show a simple "egg crate" pattern in transmission\ that is not
present at -55 dBm. With increasing rf power, screening currents begin to
slosh back and forth generating an ac magnetic field which cancels the applied
RF field. Furthermore, when these induced ac supercurrents, driven by external
microwave radiation, lock to the natural oscillations of the array there is a
range of DC currents for which there are constant frequency steps. This
current range is depicted in the number of resonances, with different
amplitudes, appearing at the same frequency.

Frequency locking is also manifested when increasing the RF power at fixed
temperature, T= 1.1 K, as seen in the transmission vs frequency plot in Fig.
4. As the RF\ power increases, the number of dips also increase. While an
increase in the current decreases the depth of the dip, it does not explain
multiple dips. We surmise that the increase in the number of dips is analogous
to Shapiro steps observed in \textit{dc }current-biased experiments with a
microwave irradiated JJ. \cite{Shapiro}\ Here, however, there is no applied
bias, only an induced dc voltage bias. Since the device is nonlinear, locking
occurs not only at $\omega_{RF},$ but \ also at harmonics of the RF\ frequency
- that is, at multiples of $\omega_{RF}$, such that \textit{n}$\omega
_{RF}=\omega_{J}$ - giving rise to discrete dips appearing at lower
frequencies and separated by 0.04 GHz as indicated by the black arrows in Fig.
4. A close-up image of a resonance splitting into two with higher RF power for
T =1.24 K is seen in the inset of Fig. 4.
\begin{figure}
[ptb]
\begin{center}
\includegraphics[
height=4.8213in,
width=3.224in
]%
{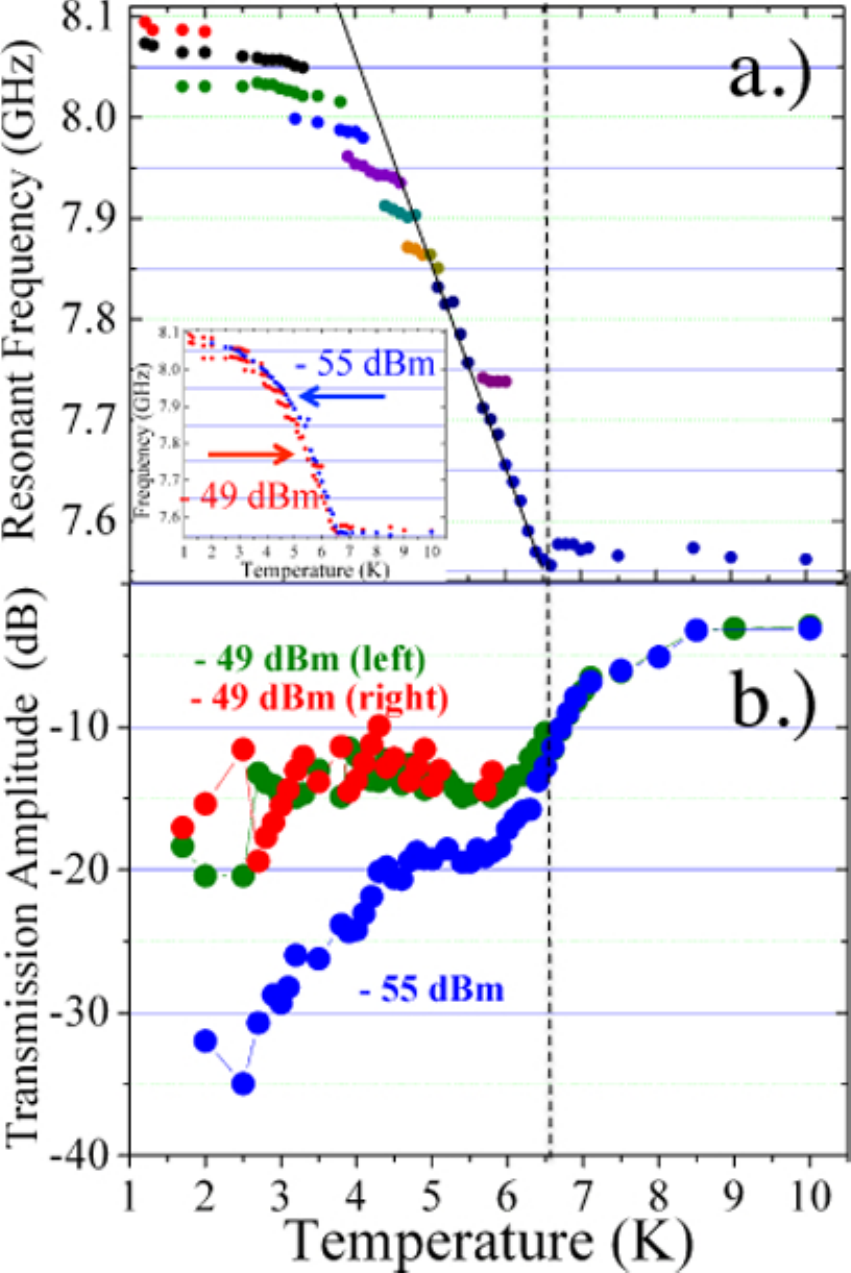}%
\caption{(Color online) a.) Resonant frequency versus temperature at - 49 dBm.
Seven steps are clearly visible for T $<$ 5 K. Each of the different colors
correspond to separate dips containing multiple resonances. A line is draw to
show the linear dependence at temperatures close to the transition temperature
of the array. Inset: Comparison between -49 dBm (red) and - 55 dBm (blue). b.)
Transmission Amplitude, S$_{21}$, versus temperature at -49 dBm (red and
green) and - 55 dBm (blue). }%
\end{center}
\end{figure}

We also analyze these results by studying how the resonant frequency and
transmission amplitude change with temperature and rf power. These results are
shown in the plots of Fig. 5. There are seven steps in resonant frequency below 5 K that are clearly observed, with each step represented by a different color, in Fig. 5 a. Considering the temperature to be proportional to the dc current and the frequency proportional to dc voltage, by the Josephson relation, $V_{DC}=n(\hbar\omega_{RF})/2e$ , then these steps look
similar to those found in I(V) curves for spontaneous emission in biased,
underdamped JJ arrays. \cite{Barbara} The range of phase locking and DC
currents in our experiments is indicated by the plateaus for each frequency;
the width of the plateau becomes smaller with higher order steps because the
current is expressed in terms of a Bessel function. \cite{newrock} Between
resonant frequencies, the phase is unlocked leaving narrow electromagnetically
opaque regions. A diagonal line is draw in the plot to show the linear
behavior between frequency and temperature at temperatures close to T$_{co}$,
the transition temperature of the array,\symbol{126}6.5 K. Comparing resonant
frequencies for -55 dBm and -49 dBm reveals two main differences: first is the
appearance of steps at -49 dBm and second, is the systematic higher shift in
the frequency vs temperature at -55 dBm below the transition temperature as
shown in the lower inset of Fig. 5 a. This shift is due to the induced rf
currents being lower.

The strength of the resonance, transmission amplitude of S$_{21}$, is dependent
on rf power as seen in Fig. 5 b. However, we find that it also varies with
temperature, in contrast to split ring resonantors with JJs \cite{anlage}, except at temperatures between \symbol{126}5 K and 6 K which
coincides to temperatures where frequency steps no longer prevail. Since the
resonance splits at -49 dBm, we label the two resonances either left as seen
in green in Fig. 5 b or right as seen in red in Fig. 5 b. The -55 dBm and -49
dBm curves crossover at \symbol{126}6.5 K which is the same temperature that
the resonant frequency flattens out as seen in Fig. 5 a. This is indicated by
a vertical dash line in Fig. 5. There is still a resonance beyond 6.5 K that
doesn't change in frequency suggesting superconducting quasiparticles
tunneling until T \symbol{126} 8.5 K.

In conclusion, our results demonstrate that a wireless JJ array in a waveguide
can controllably switch on and off a microwave signal with temperature and
input power. The array can also modulate the signal over a range of
frequencies and amplitudes by varying either input power, temperature, or dc
magnetic field (not shown here). Modulating signals with temperature is
traditionally considered to be quite slow; however, for JJ arrays, the
response occurs rapidly and without noticeable delay. The signal is sharp and
robust without processing the data and does not require the aid of amplifiers
and filters in the experimental configuration. Moreover, our results reveal
qualitatively new phenomena for tunable metamaterials that occur as the input
power increases, breaking the central resonance down to discrete frequencies.
Finally, this alternative perspective on metamaterials may provide new strategies for
future metamaterial design and tunability.

Acknowledgements. None of this work would have been possible without the
generosity of Steve Anlage and Allen Goldman. The author thanks both of them
for this opportunity. Beneficial discussions with Chris Lobb, Donhee Ham and
Bert Halperin are also gratefully acknowledged. This work was supported by the
Intelligence Community Postdoctoral program and carried out at the University
of Maryland, Center for Nanophysics and Advanced Materials, College Park, MD.

*Email address: lladams@seas.harvard.edu

\bigskip


\begin{thebibliography}{99}
\bibitem {squid}Superconducting transition temperature was measured with a
Quantum Design SQUID\ magnetometer.
\end{thebibliography}
\end{document}